\documentclass[prl,twocolumn,showpacs,preprintnumbers,amsmath,amssymb]{revtex4-2}
\usepackage{graphicx}
\usepackage{dcolumn}
\usepackage{bm}
\begin{document}
\title{\bf Two New Members of the Covalent Organic Frameworks Family: Crystalline 2D-Oxocarbon and 3D-Borocarbon Structures}
\author{N. Hassani$^{1}$, A. Movafegh-Ghadirli$^{2}$, Z. Mahdavifar$^{3}$, F. M. Peeters$^{4}$, and M. Neek-Amal$^{2,4}$}
\affiliation{
$^1$ Department of Physical Chemistry, Faculty of Chemistry, Razi University, 67144-14971 Taq-e Bostan, Kermanshah, Iran.\\
$^2$Department of Physics, Shahid Rajaei University,
Lavizan, Tehran 16785-136, Iran.\\
$^3$Department of Chemistry, Faculty of Science, Shahid Chamran University of Ahvaz, Ahvaz, Iran.\\
$^4$Departement Fysica,
Universiteit Antwerpen, Groenenborgerlaan 171, B-2020 Antwerpen,
 Belgium.}
\date{\today}

\begin{abstract}

While graphene oxide (GO) is representative of a disordered phase of oxocarbons with lackluster electronic properties, the coexistence of ordered, stoichiometric solid-state carbon oxides with graphene brings renewed momentum to the exploration of two-dimensional crystalline oxocarbons. This enduring subject, spanning decades, has recently witnessed significant advancements. In this context, our study delves into a novel material class, COF-66, notable for its meticulously ordered two-dimensional crystalline structure and intrinsic porosity. Employing global optimization algorithm alongside density-functional calculations, our investigation highlights a standout member within the COF-66 family—an exceptional quasi-flat oxocarbon ($C_6O_6$)—exhibiting an unconventional oxygen-decorated pore configuration. This pioneering study introduces $C_6O_6$ as an innovative entrant into the crystalline carbon oxide arena, augmenting the established understanding alongside the well-recognized graphene oxide and two graphene monoxide, i.e. a-GMO and b-GMO. Expanding the exploration, the COF-66 series encompasses 2D-porous carbon nitride (C$_6N_6$) and the recently synthesized 2D-porous boroxine ($B_6O_6$), adhering to a generalized stoichiometry of $X_6Y_6$, where X = B, C, and Y = B, N, O, with X $\neq$ Y. Remarkably, the entire COF-66 ensemble adopts a 2D-crystalline framework, with the exception of $C_6B_6$, which assumes a distinct 3D-crystalline arrangement. Furthermore, we explore the hydrogen storage capability of $C_6B_6$.
Employing the PBE (HSE06) level of theory, our electronic structure calculations yield band gap values of 0.01 (0.05)\,eV, 3.68 (5.29)\,eV, 0.00 (0.23)\,eV, and 1.53 (3.09)\,eV for $B_6N_6$, $B_6O_6$, $C_6B_6$, and $C_6N_6$, respectively, reinforcing and aligning with prior investigations. Noteworthy is the introduction of spin-orbit coupling (SOC), leading to the emergence of a 0.84\,eV band gap in $B_6N_6$. Our comprehensive study extends beyond electronic properties, encompassing mechanical attributes, including the three-dimensional counterpart, $C_6B_6$, while simultaneously revealing captivating magnetic and optoelectronic phenomena within $B_6N_6$.
\end{abstract}
\maketitle

\begin{figure}
\centering\includegraphics[width=0.90\linewidth]{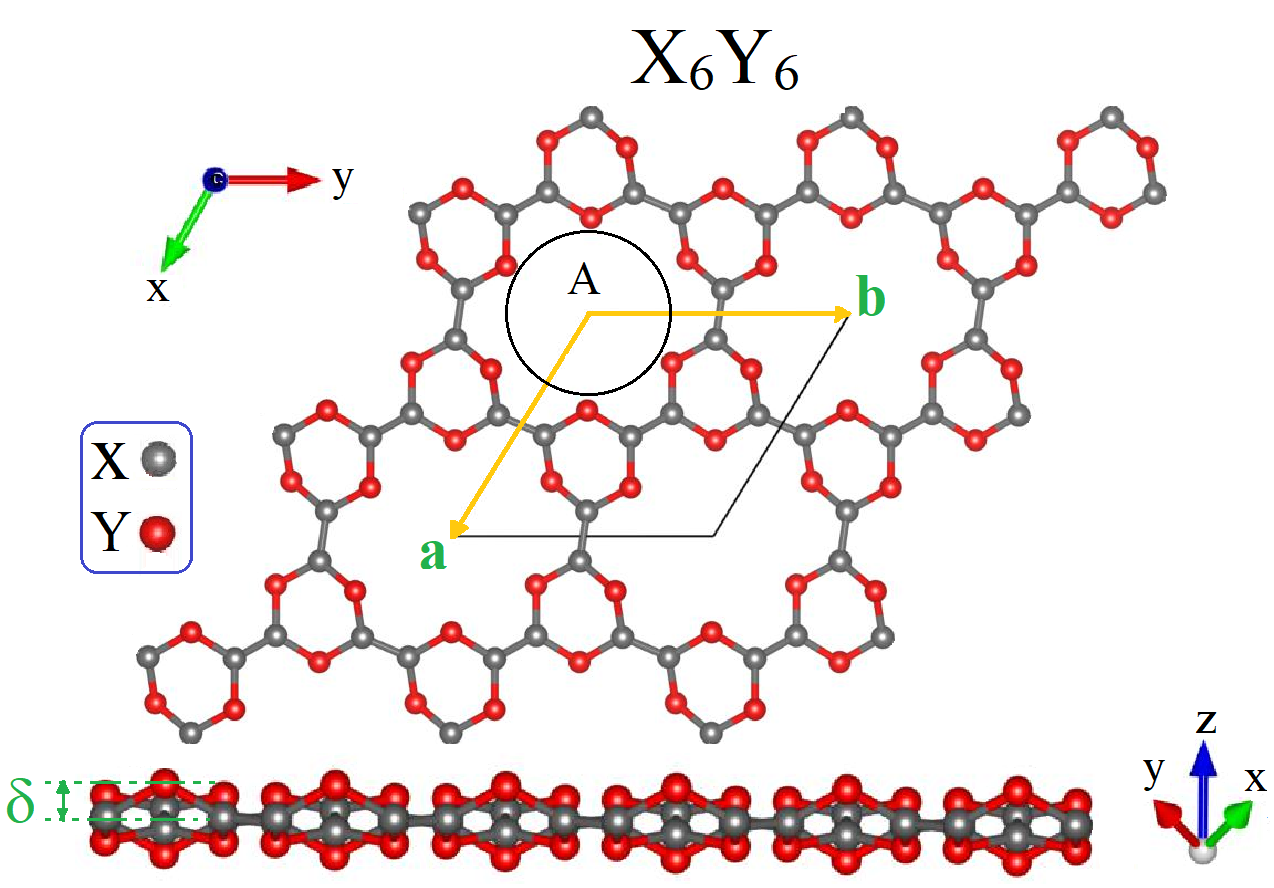}
\caption{Top and side view of $X_6Y_6$ lattice structure (for this image $X=C$ and $Y=O$).
$\delta$ indicates the out-of-plane position of O atoms with respect to the plane of C atoms. Primitive cell borders: black;
 primitive lattice vectors: yellow. Color code of balls: C: gray; O: red.}
\label{Fig1}
\end{figure}

\section{INTRODUCTION}
Carbon dioxide ($CO_2$), carbon monoxide ($CO$), carbon suboxide ($C_3O_2$), mellitic anhydride ($C_{12}O_9$), and 1,3,5-trioxane ($C_3H_6O_3$) are among the known oxocarbons\,\cite{de2016oxocarbons}. Other oxocarbons that are either unstable or metastable include cyclic radialene-type oxocarbons, $C_nO_m$ with $n$= 4, 5, and 6\,\cite{seitz1992oxocarbons, hassani2020co}. However, their anions are stable\,\cite{schroder1999mass}. In addition to the well-known properties of $CO_2$ and $CO$, carbon suboxide spontaneously polymerizes into a red, yellow, or black solid\,\cite{ellern2001structure}, while 1,3,5-trioxane is extensively used in the production of polyoxymethylene plastics that exhibit excellent dimensional stability, for which Hermann Staudinger was awarded the 1953 Nobel Prize in Chemistry.

 Graphene oxide (GO) is another widely recognized oxocarbon that belongs to the graphene family, with a 2D aromatic scaffold composed of carbon atoms with hybridization of $sp^2$ and $sp^3$ that are associated with in-plane and tetrahedral structures, respectively. GO contains various functional groups, such as epoxide, carbonyl, lactone, carboxyl, and hydroxyl, which are randomly located at the edge of the basal plane. The number and arrangement of these functional groups are challenging to control during the GO synthesis process\,\cite{luo2009photoluminescence}.
About a decade ago, an ordered crystalline graphene-based oxocarbon, a-GMO, with D$_{2h}$ symmetry and a quasi-hexagonal unit cell was synthesized by annealing multilayered GO with precisely determined O:C stoichiometry of 1:1\,\cite{mattson2011evidence}. a-GMO is a semiconductor with an indirect band gap of 0.6\,eV\,\cite{pu2013strain}. More recently, Radevych \emph{et al.} proposed another phase of GMO, b-GMO, with symmetry group D$_{6h}$ using selected area electron diffraction, high-resolution transmission electron microscopy, and density functional theory (DFT) calculations\,\cite{radevych2022beta}. The chemical formulas of a-GMO and b-GMO are $C_2O_2$ and $C_6O_6$, respectively, with oxygen atoms arranged in a specific order on the C-C bonds of the graphene lattice.

Furthermore, in recent years, several new covalent organic frameworks (COFs) have been synthesized beyond the two pioneering COFs, COF-1, and COF-5 \cite{cote2005porous}. Among these new COFs are 2D-porous graphitic carbon nitride~\cite{bafekry2020two} (here g-$C_6N_6$) and 2D-porous boroxine ($B_6O_6$) \cite{cao2004synthesis, stredansky2018surface}, which are isostructural and isoelectronic 2D materials with hexagonal lattice structures. For uniformity, we will use the notations $C_6N_6$ and $B_6O_6$ to refer to these two structures.
COFs are synthesized using various methods and different linkages \cite{zhao2017covalent}. Due to the periodic voids in their structure, they can act as guest-templated host networks that can host active molecules/atoms in regular nanoarrays \cite{plas2016nanopatterning, khan2021supramolecular, furukawa2009storage,neek2012strain, el2007designed}. This unique property makes COFs attractive for various applications, including nanopatterning, supramolecular chemistry, and gas storage. Additionally, COFs can be functionalized with different functional groups to enhance their host-guest properties and make them suitable for specific applications.
For example, the $C_6N_6$ monolayer has been shown to be an effective adsorbent for CH$_4$O, 2-pentanone, and limonene, with adsorption energies of 0.55\,eV, 0.68\,eV, and 0.81\,eV, respectively \cite{zhang2021g}. The $C_6N_6$/SiP-GaS heterojunction is a type-II heterojunction, with its band edge straddling the redox potential of water splitting \cite{yang2022two}.
In addition, $B_6O_6$ deposited on an Au(111) surface has excellent tunability of its optical properties from UV to visible and is suitable for adsorbing alkalis. The adsorption energy of potassium is 0.52 and 0.61\,eV larger than those of sodium and Lithium, respectively \cite{ullah2019theoretical}

The successful synthesis of 2D COFs ($B_6O_6$ and $C_6N_6$) has motivated us to search for a broader family of porous COFs comprising B, C, N, and O elements with the same symmetry as $B_6O_6$ and $C_6N_6$. To discover potential $X_6Y_6$ structures (COF-66 family) with X = B, C, and Y = B, N, O; where X is not equal to Y, we employed an evolutionary algorithm. This approach has been proven to be the most successful technique to search for global minima and is capable of predicting unknown crystal structures \cite{oganov2011evolutionary,oganov2011modern}.

  In our study, we employed dispersion-corrected hybrid DFT methods and ab-initio molecular dynamics simulations to post-process our predicted COFs. This enabled us to characterize various properties of the COFs, including the crystalline two-dimensional oxocarbon $C_6O_6$. To validate our results, we compared them with those of other crystalline two-dimensional oxocarbons, such as a-GMO and b-GMO \cite{radevych2022beta,mattson2011evidence}. While there have been some older reports on the polymerization of carbon suboxide\,\cite{snow1978poly,carofiglio1986carbon}, no theoretical or experimental studies have reported crystalline oxocarbon $C_6O_6$, whether it be freestanding or deposited on a substrate. Moreover, in recent years, machine-guided methods have become increasingly popular for discovering new crystalline materials \cite{oganov2019structure} and for the self-assembly of nanopatterned particles/polymers \cite{pula2022solvent}. Here, we also used the evolutionary algorithm-based crystal structure prediction~\cite{oganov2006crystal} and implemented it in the USPEX code \cite{popov2021novel}, which is also suitable for predicting two-dimensional materials.

\begin{table*}[ht]
\centering
 \caption{The electronic structure (ES): the calculated lattice constants for $X_6Y_6$,
  $a$ and $b$ (in \,\AA{}); the electronic band gap at HSE06 level for a 1$\times$1 unit cell, $\Delta{^\prime}$ (in eV); and the out-of-plane height ($\delta$ in \,\AA{}), the phonon band gap, $\Delta_p$ (in cm$^{-1}$); the linkage between successive hexagons, and the X-X and X-Y bond lengths (in \AA).}
\scalebox{1.00}{
 \begin{tabular}{|c|c|c|cccccc|}
 \hline
 X, Y &ES&$(a,b)$&$\Delta{^\prime}$(/SOC)&$\delta$&$\Delta_p$&$Linkage$&X-X&X-Y\\
 \hline
  C, O& Insulator      & (6.93,6.93) &4.21 & 0.64&156.88& C-C&1.35&1.3(7) \\
  B, N &Semiconductor           & (8.23,8.23)   &$~$0.05(0.84)&0.0 & 102.93& B-B&1.71&1.4(1)  \\
    B, O &Insulator      & (7.81,7.81)  &5.29 &0.0 &221.34& B-B &1.72&1.3(8) \\
  C, N &Semiconductor   & (7.14,7.14)  & 3.09  &0.0&181.54 & C-C&1.51&1.3(5)  \\
  C, B &Semiconductor          & (8.61, 9.98)&0.23 &-& 25.67 &- &1.48&1.6(1)  \\
  N, O &Gas             & - & -  & - &-& -&-&-\\
  \hline
 \end{tabular}
 }
 \label{Table1}
 \end{table*}
\begin{figure*}
\centering\includegraphics[width=0.850\linewidth]{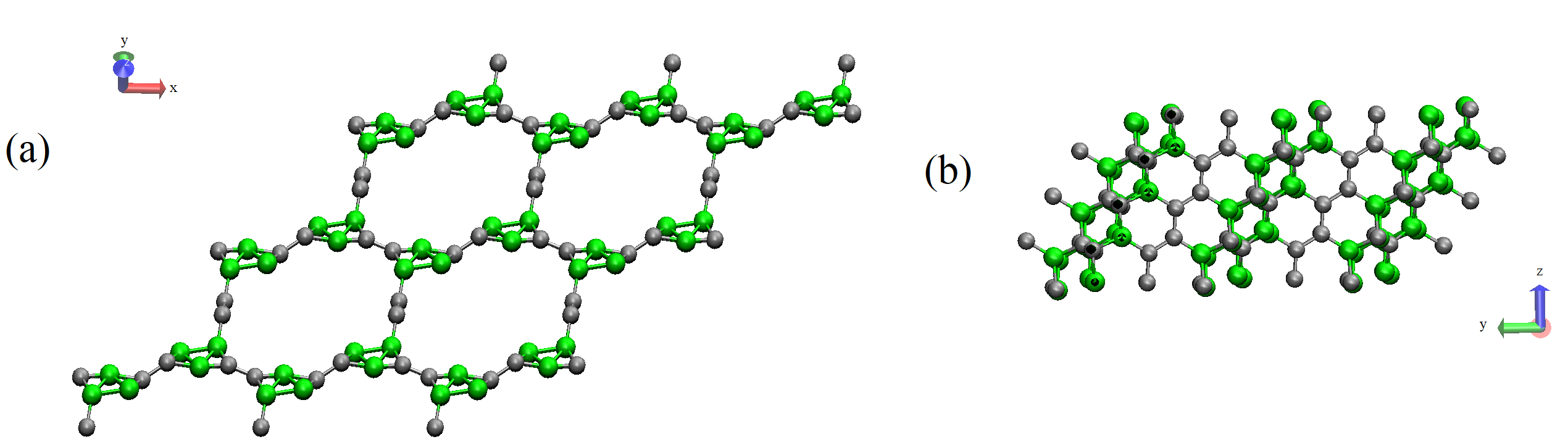}
\caption{Two different views of the 3D-$C_6B_6$ structure are presented to showcase its porosity and stiffness. (a) The left panel provides a view of the porous side, highlighting the structural voids and potential for gas adsorption. (b) The right panel, on the other hand, shows the dense side, emphasizing the high structural stability of  3D-$C_6B_6$. Color code of balls: C: gray; B: green.}
\label{Fig2}
\end{figure*}

\begin{figure*}
\centering\includegraphics[width=0.85\linewidth]{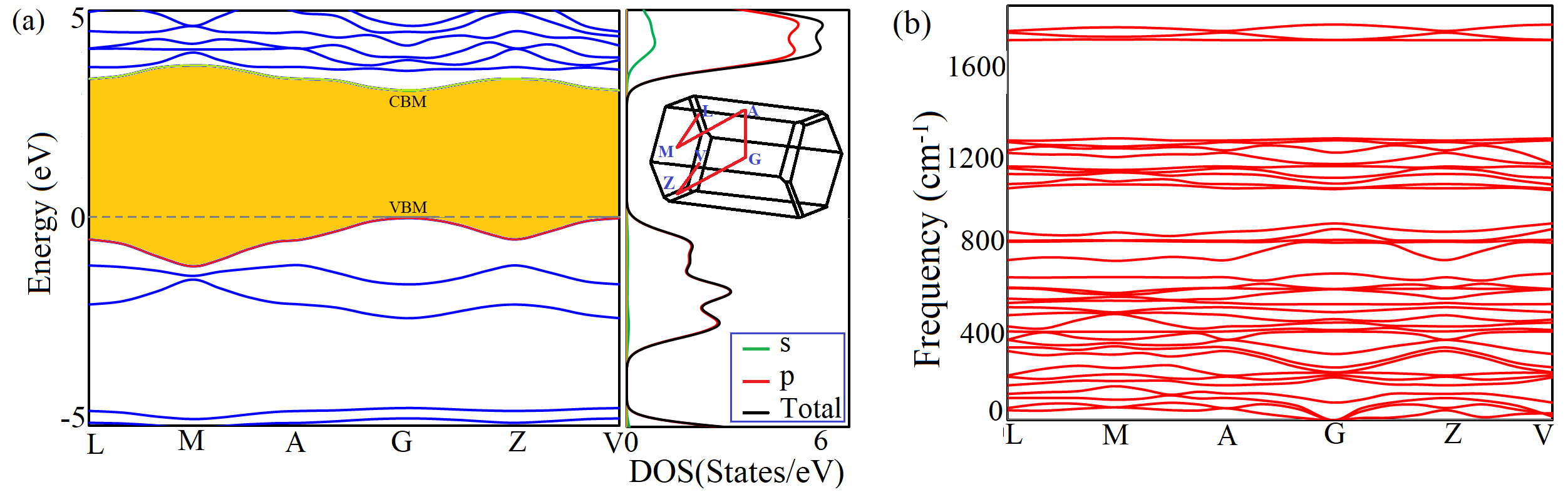}
\caption{(a) The electronic band structure (left) along with the density of states (right) of
$C_6O_6$. The inset in DOS shows the symmetry lines and important points in the first Brillouin zone (FBZ). (b) The phonon band structure of $C_6O_6$ along the high symmetry point of  FBZ (shown in the inset).}
\label{Fig3}
\end{figure*}
\section{Computational details}
\subsection{Global minimum search}
i) The ground-state structure determination for two COF-66 members, namely 2D-$C_6O_6$ and 2D-$C_6N_6$, employed the global optimization technique implemented in the USPEX code \cite{oganov2006crystal,lyakhov2013new}. To ensure global minima were achieved, we generated 4500 initial structures across 30 generations. Selecting energetically favorable isomers with relative energies within the [0-1.5]\,eV range compared to the most stable structure, we pursued the global minimum structure within the predicted set. This process involved meticulous re-optimization of the structures to yield the accurate lowest energy 2D-$C_6O_6$ crystal structure. To maintain the focus on 2D crystalline structures, we excluded molecular/polymeric candidates. Ultimately, the lowest energy 2D structure, $C_6O_6$, was identified. Another circular 2D oxocarbon structure was predicted but omitted due to the presence of imaginary frequency phonons.

The structure of $C_6N_6$, a known system, was discovered utilizing the USPEX code. In parallel, the code generated additional crystalline graphitic nitride structures, including g-$C_3N_4$ \cite{liu2016graphitic}. However, as our interest lay exclusively in isostructures of the type $X_6Y_6$, these other structures were not considered within this study.

ii) Despite an extensive 30-generation search, the USPEX code did not successfully predict the structure of $B_6N_6$. To obtain this structure, we substituted a carbon atom in $C_6N_6$ with a boron atom.

iii) The evolutionary structure search failed to predict the 3D-$C_6B_6$ structure, prompting us to construct it through a sequence of substitutions. This involved replacing B atoms with N in $C_6N_6$ and subsequently performing diverse 2D and 3D energy minimizations employing DFT. The resulting 3D-$C_6B_6$ structure, characterized by porosity and remarkable mechanical stability, was validated through positive frequency confirmation (further details are provided below on the structural properties).

iv) While the USPEX code did not predict the $B_6O_6$ structure across 30 generations of searching, we succeeded in obtaining this structure by substituting a carbon (nitrogen) atom in $C_6N_6$ with a boron (oxygen) atom.

It is pertinent to note that, although the USPEX code could potentially explore alternative structures, we made informed structural decisions based on the available information from $C_6N_6$ and $B_6O_6$ structures, thus bypassing the need for additional USPEX iterations. Therefore, we optimized the predicted COF-66 structures using DFT calculations and found them to be stable. It is interesting to observe that $B_6N_6$, $B_6O_6$, and $C_6N_6$ are flat structures, while $C_6O_6$ has a small out-of-plane height ($\delta=0.64$\,\AA).~ Table I summarizes some of the structural and electronic properties of COF-66, such as the electronic band gap at the HSE06 level ($\Delta^{\prime}$), the out-of-plane height ($\delta$), the phonon band gap ($\Delta_p$), the linkage between successive hexagons, and the X-X bond lengths and some of the X-Y bond lengths.

\begin{figure*}[tbp]
\centering\includegraphics[width=1.00\linewidth]{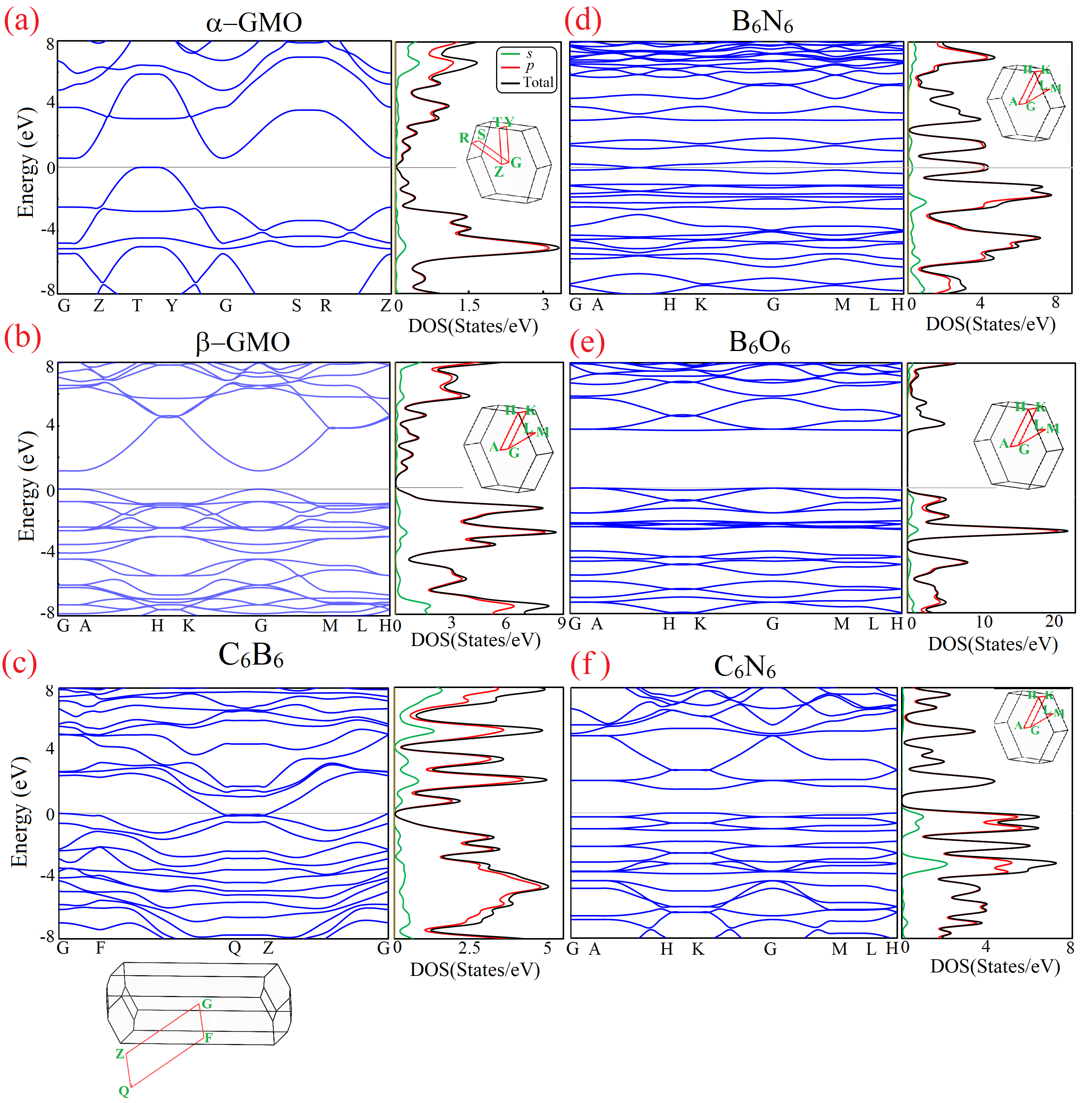}
\caption{In each panel: The electronic band structure (left) along with density of states (right) of
(a) a-GMO, (b) b-GMO, (c) $C_6B_6$, (d) $B_6N_6$, (e) $B_6O_6$, and (f) $C_6N_6$ systems.
The inset in PDOS shows the symmetry lines and points in the Brillouin zone.}
\label{FigS2}
\end{figure*}

\begin{figure}[tbp]
\centering\includegraphics[width=0.95\linewidth]{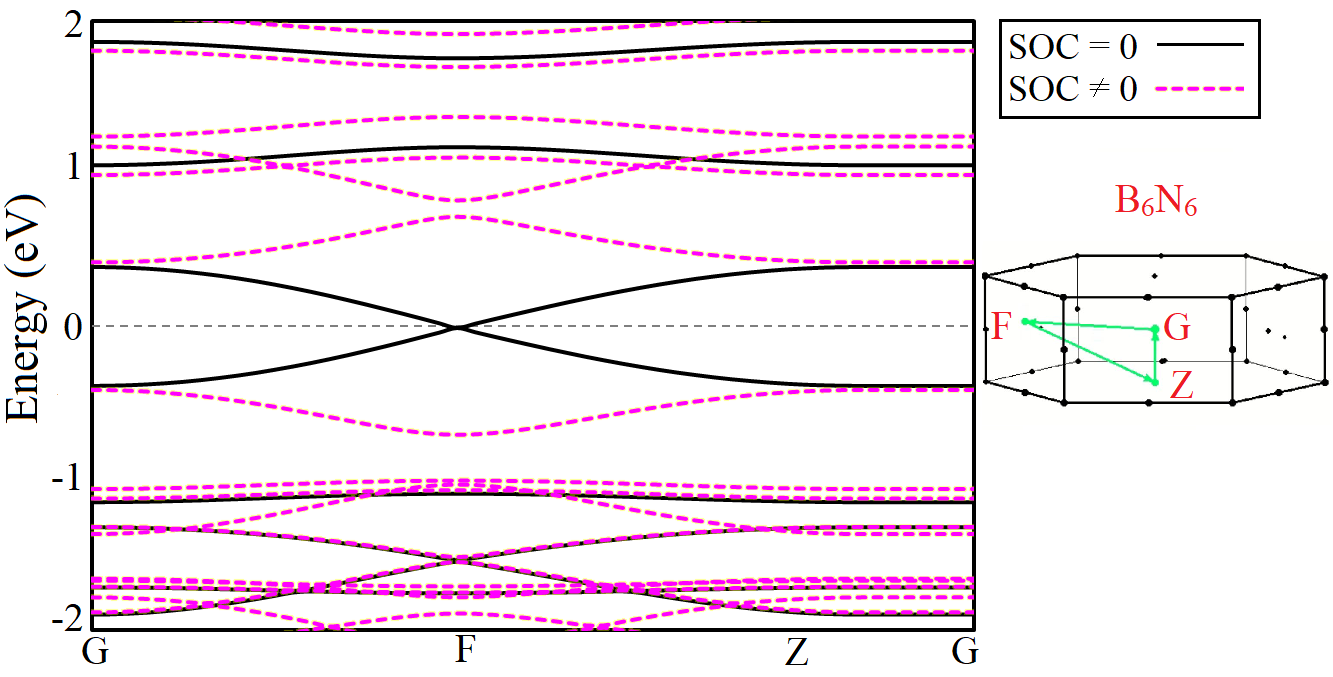}
\caption{The electronic band structure of $B_6N_6$ with (dashed magenta curves) and without (black curves) inclusion of spin-orbit coupling  along the high symmetry point of  FBZ (shown in the inset).}
\label{Fig4}
\end{figure}

\begin{figure*}[tbp]
\centering\includegraphics[width=1.00\linewidth]{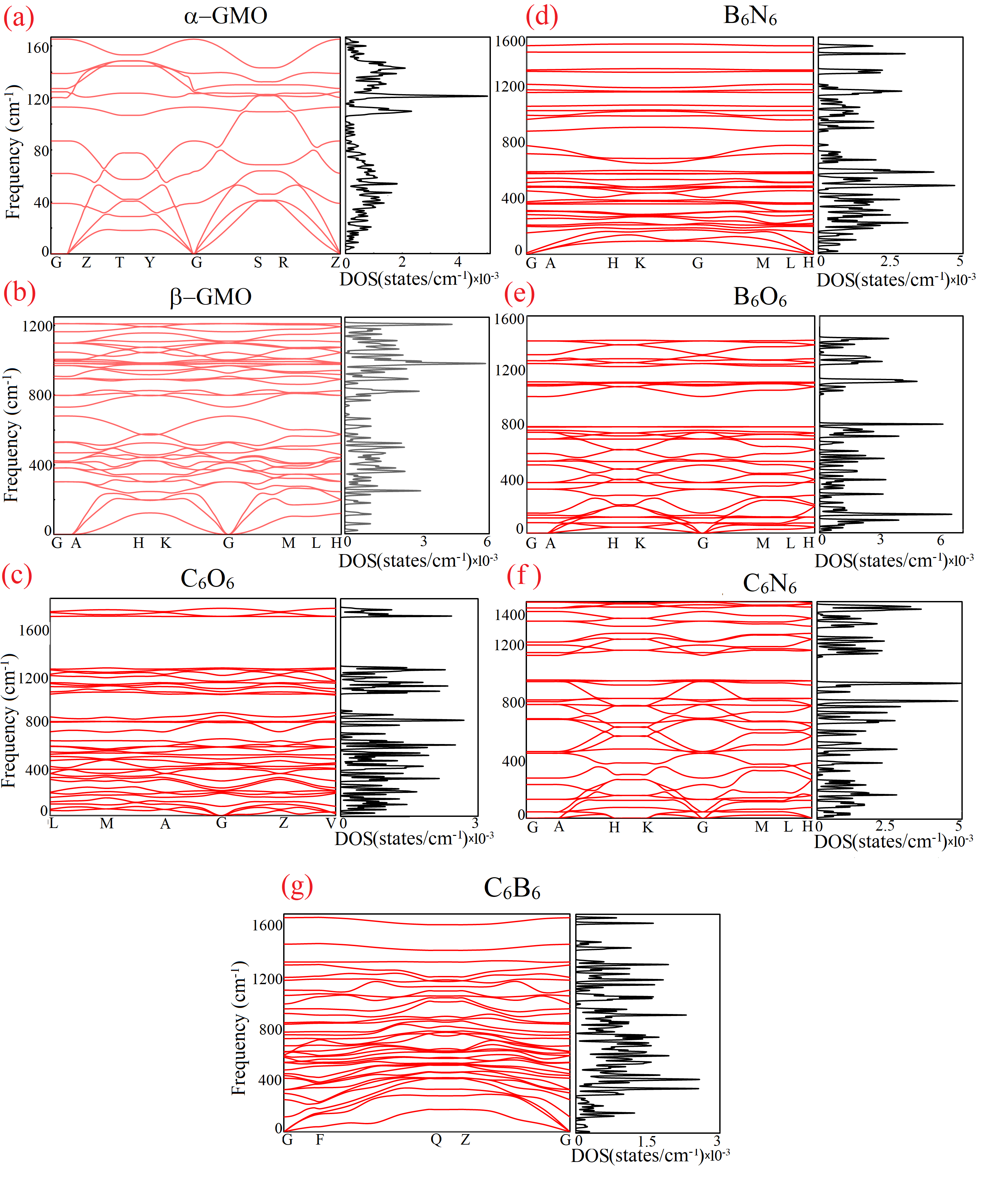}
\caption{In each panel: The phonon band structure (left) along with phonon density of states (right) of
(a) a-GMO, (b) b-GMO, (c) $C_6B_6$, (d) $B_6N_6$, (e) $B_6O_6$, (f) $C_6N_6$, and (g) $C_6B_6$ systems.
}
\label{FigS4}
\end{figure*}

\subsection{Ab initio calculations}

The CASTEP code, as implemented within the Material Studio software [27], has been employed to perform comprehensive atomic geometry optimizations for the predicted COF-66 and two distinct GMOs (a-GMO and b-GMO). In our study, we harnessed two prominent functionals in the realm of density functional theory: the HSE06 (Heyd-Scuseria-Ernzerhof) hybrid functional and the Perdew-Burke-Ernzerhof (PBE) functional \,\cite{heyd2003hybrid,perdew1996generalized}. Specifically, the HSE06 functional, notable for its suitability in systems featuring significant electron correlation effects, has been incorporated. By integrating a portion of Hartree-Fock exchange into conventional DFT exchange-correlation functionals like PBE, the HSE06 approach notably enhances the precision of energy band gaps, electronic structures, and material properties compared to purely DFT-based computations. Its efficacy is particularly pronounced in describing systems characterized by robust electronic interactions, thus establishing its reliability in the context of our electronic structure investigations.

To capture van der Waals interactions, we applied Grimme’s empirical dispersion-corrected method\,\cite{grimme2006semiempirical}. The relaxation procedure has been conducted until energy and force variations reached 0.01\,meV and 0.01\,eV\,\AA$^{-1}$,~ respectively. To exclude any spurious interactions arising from adjacent layers, a vacuum spacing of 15\,\AA, along the $z$-axis was incorporated.

For orbital calculations encompassing the determination of highest occupied crystal orbitals (HOCO), lowest unoccupied crystal orbitals (LUCO), electron density, and Fukui functions\,\cite{delley2000molecules}, the DMol3 package was employed.

Furthermore, the investigation extended to ab-initio molecular dynamics (AIMD) simulations for both 1$\times$1 and 4$\times$4 unit cells utilizing the CASTEP code. These simulations were conducted under the NVT ensemble at 297\,K for a duration of 10\,ps, with supplementary movies providing an illustrative depiction of the outcomes.

To meticulously explore the impact of spin-orbit coupling (SOC) on the electronic structure calculation of B6N6, we adopted a full relativistic pseudopotential approach using the Quantum ESPRESSO software package\,\cite{giannozzi2009quantum}. SOC, a fundamental relativistic phenomenon that intertwines the electron’s spin and orbital angular momentum, emerges as a crucial factor in dictating the electronic behavior of materials containing heavy elements or exhibiting strong spin-orbit interactions. Within the framework of DFT, we have thoughtfully enabled SOC in our computations, allowing for the precise capture of spin-split energy bands and a comprehensive understanding of its implications for the electronic properties of  $B_6N_6$.

\section{Results and discussion}
\subsection{Structural and electronic properties of a-GMO, b-GMO, and COF-66.}
 The fully relaxed structure of $C_6O_6$ is shown in Fig.~\ref{Fig1}, and the other members of COF-66 are shown in Fig.~S1 in the Supporting Information (SI). The lattice constants ($a$ and $b$) and other physical properties of COFs are given in Table I.

In $C_6O_6$, both C and O atoms are embedded in a hexagon containing 3 C and 3 O with a symmetry group of C2/M. The symmetry groups of $C_6N_6$ ($B_6O_6$) and $B_6N_6$ ($C_6B_6$) are P6/MMM and C2/M, respectively. $C_6N_6$, $B_6N_6$, and $B_6O_6$ are completely planar, while $C_6B_6$ has a 3D-lattice structure. The out-of-plane height for O atoms is found to be $\pm$\,1.04\,\AA{}, $\pm$\,1.03\,\AA{}, and $\pm$,0.32,\AA{} from the carbon plane in the a-GMO, b-GMO, and $C_6O_6$ systems, respectively. However, the C$-$C bond lengths of the a-GMO (b-GMO) and $C_6O_6$ are 1.58\,\AA{} (1.60\,\AA{}) and 1.38\,\AA{}, respectively, which are close but smaller than typical sp$^3$ bond lengths (1.54\,\AA{}). The obvious difference in the lattice structure of GMOs and COF-66 is the presence of a hole in the vicinity of COF-66 cells.

Surprisingly, rather large values for the Young, shear, and bulk moduli and the Poisson ratio of 3D-$C_6B_6$ were found to be 236\,GPa, 99\,GPa, 127\,GPa, and   0.2, respectively. The latter shows that notwithstanding the large porosity of the system, 3D-$C_6B_6$ is one of the stiffest porous materials ever reported \cite{Kovacik99}. Recently, tensile measurements were conducted on a COF nano-film with a thickness of 85 nm, which resulted in Young's modulus of 37 $\pm$ 15 GPa\,\cite{Missale2022}. Note that due to the smaller pore size of 3D-$C_6B_6$  compared to the pores of other COFs \cite{furukawa2009storage,el2007designed}, its density (1.66\,gcm$^{-3}$) is higher than that of COF-102, COF-105, and COF-108~\cite{el2007designed}. Therefore, the potential applications of 3D-$C_6B_6$ (and in general COF-66) would differ from those reported for COF-1, COF-5, COF-6, COF-8, and COF-10, which are considered among the most porous and effective adsorbents for hydrogen, methane, and carbon dioxide in gas storage applications~\cite{furukawa2009storage}. However, due to the smaller pore size, hydrogen storage/selectivity and water splitting applications for COF-66, which are typically expected from 2D membranes with atomic apertures\cite{thiruraman2020gas}, could be explored. In Fig.~\ref{Fig2} the left panel offers a view of the porous side, the structural voids, and the potential for gas adsorption.~In contrast, the right panel depicts the dense side, underscoring the high structural stability of the 3D-$C_6B_6$ material. Together, these perspectives present a comprehensive overview of the distinctive properties of $C_6B_6$. Future studies are needed to investigate gas separation using a scaled-up version of COF-66.

The electronic band structure and density of states (DOS) of $C_6O_6$ is shown in Fig.~\ref{Fig3}  and for other $X_6Y_6$ and a-GMO, b-GMO are shown in Fig.~\ref{FigS2}. The band gaps ($\Delta$) of a-GMO, b-GMO, and $C_6O_6$ at the PBE level of theory are found to be 0.58 eV, 1.15 eV, and 3.08 eV, respectively. Since PBE tends to underestimate $\Delta$, we also calculated the band gap ($\Delta{^\prime}$) using the HSE06 hybrid functional. These results indicate that a-GMO, b-GMO, $C_6N_6$, $B_6N_6$ and $C_6B_6$ exhibit semiconducting and   $C_6O_6$ and $B_6O_6$ exhibit insulating behavior. The latter refers to the lower conductivity of $C_6O_6$, where the population of electronic states in a-GMO and b-GMO near the Fermi level is larger than in $C_6O_6$. Both conduction band minima (CBM) and valence band maxima (VBM) lie at the gamma point for b-GMO, $C_6O_6$, and $B_6O_6$ (for $C_6N_6$ at the K-point), while in a-GMO, CBM and VBM lie at the $\Gamma$ and T points, respectively. As a result, b-GMO ($C_6O_6$, $C_6N_6$, $B_6O_6$) and a-GMO have a direct and indirect band gap, respectively.

At the PBE (HSE06) level of theory, the band gap values for $B_6N_6$, $B_6O_6$, $C_6B_6$, and $C_6N_6$ are 0.01 (0.05)\,eV, 3.68 (5.29)\,eV, 0.00 (0.23)\,eV, and 1.53 (3.09)\,eV, respectively. Our calculated band gap values for $C_6N_6$ and $B_6O_6$ are in good agreement with those reported in Refs.~\cite{srinivasu2014porous,ullah2019theoretical}.
It's worth noting that a band gap can be opened in $B_6N_6$ by including spin-orbit coupling (SOC) in the spectrum (see Fig.~\ref{Fig4}), resulting in a band gap of 0.84\,eV. This value is smaller than the reported band gap in Ref.~[36], which may be due to a not well-optimized structure in that study.
It has also been found that $B_6N_6$ exhibits interesting magnetic and optoelectronic properties \cite{abdullahi2021antiferromagnetic}.

Furthermore, for all COF-66 compounds, the CBM and VBM are primarily contributed by electrons in the $p$ orbitals. This provides valuable information on the electronic structure of these compounds, which can be useful for designing materials with specific electronic properties.

It's worth noting that, unlike the other compounds studied, $C_6O_6$ exhibits three isolated bands below the Fermi level. This unique electronic structure could have interesting implications for the material's properties and potential applications. To assess the kinetic stability of GMOs and COF-66, we calculated the phonon band structure and phonon density of states at T = 0\,K, as shown in Fig.~\ref{FigS4}. Due to the presence of real phonon frequencies, it is anticipated that the relaxed structures of GMO and COF-66 will exhibit stability. Our AIMD simulations support this prediction, showing that COF-66 remains thermally stable at room temperature (298\,K).

Fig.~S2 in the Supplementary Information illustrates the frontier crystal orbitals of the GMOs and COF-66 structures, specifically the highest occupied crystal orbital (HOCO) and lowest unoccupied crystal orbital (LUCO).
In COF-66, the HOCO and LUCO on the atoms near the pore mouth protrude more than those on other atoms, indicating that electron acceptor/donor molecules are more likely to be adsorbed on the atoms at the pore mouth of COF-66.

\begin{figure}
\centering\includegraphics[width=1.00\linewidth]{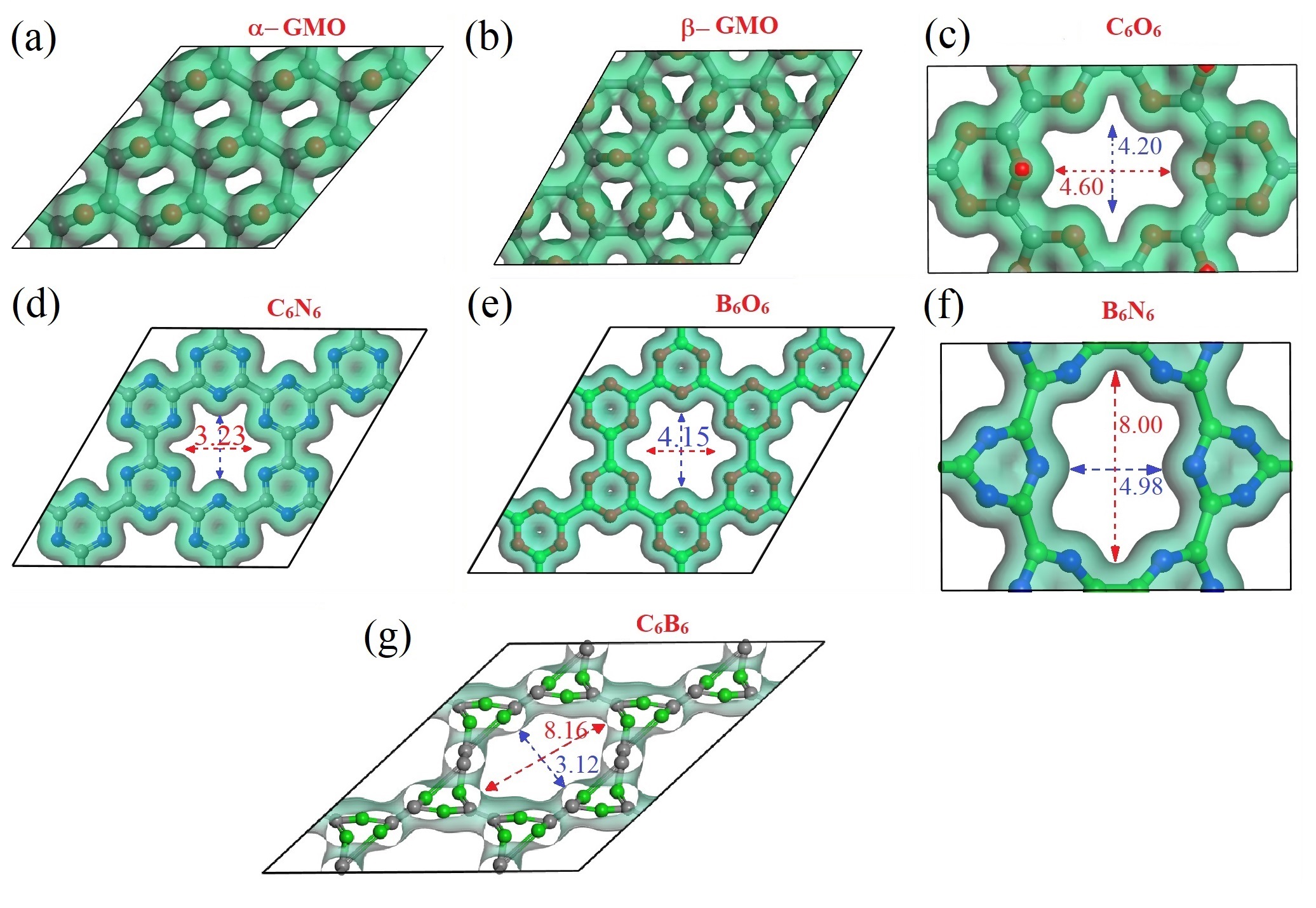}
\caption{Electron density isosurface of (a) a-GMO, (b) b-GMO, (c) $C_6O_6$, (d) $C_6N_6$,
(e) $B_6O_6$, (f) $B_6N_6$, and (g) $C_6B_6$ with the isovalue of 0.02 $e$/\AA{}$^3$.
Color code of balls: C: gray; O: red; B: green; N: blue. The dimension of pores is shown by arrows. }
\label{Fig5}
\end{figure}

Fig.~\ref{Fig5} shows the electron density isosurface of both GMOs and COF-66. The electron density at the center of the hexagons in GMOs is smaller than in other regions. The hole within the $C_6O_6$ lattice has an oval shape with dimensions of 6.40\,\AA{}$\times$\,4.20\,\AA{} and an area of approximately 21\,\AA$^2$. This size is smaller than that of $B_6O_6$ (23\,\AA$^2$)\,\cite{ullah2019theoretical}. The dimensions of the hole for $C_6B_6$ and $B_6N_6$ are 8.16\,\AA{},$\times$\,3.12\,\AA{}and 8.00\,\AA{}$\times$4.98\,\AA,~respectively. The hole of $C_6O_6$ ($C_6N_6$, $B_6O_6$) is circular, while that of $B_6N_6$ ($C_6B_6$) is elliptical.

\begin{figure}[tbp]
\includegraphics[width=1.0 \linewidth]{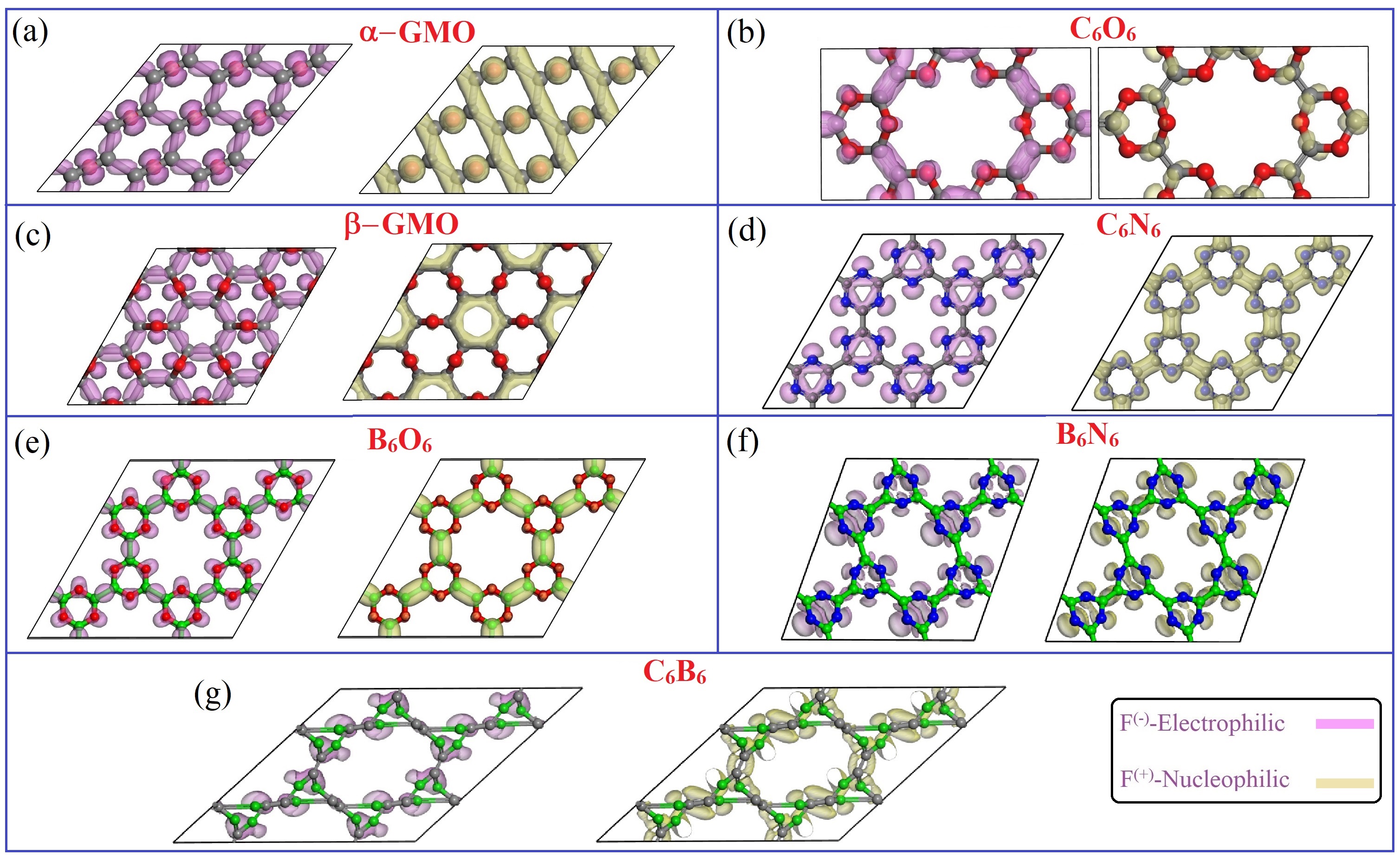}
\caption{The electrophilic Fukui functions (pink isosurfaces) and the nucleophilic  Fukui functions
(yellow isosurfaces) of (a) a-GMO, (b) $C_6O_6$, (c) a-GMO, (d) $C_6N_6$,
(e) $B_6O_6$, (f) $B_6N_6$, and (g) $C_6B_6$ at isovalue of 0.003 a.u.
Color code of balls: C: gray; O:red; B:green; N: blue.}
\label{fig6}
\end{figure}

Fig.~\ref{fig6} shows that, based on Fukui function analysis, atoms at the pore mouth of $C_6O_6$, $C_6B_6$, and $B_6N_6$ are susceptible to adsorption of nucleophilic species, while carbon (nitrogen) atoms are capable of attracting electrophilic species.

\subsection{Thermodynamic properties of the $X_6Y_6$ structures}
\begin{figure}[tbp]
\includegraphics[width=1.0 \linewidth]{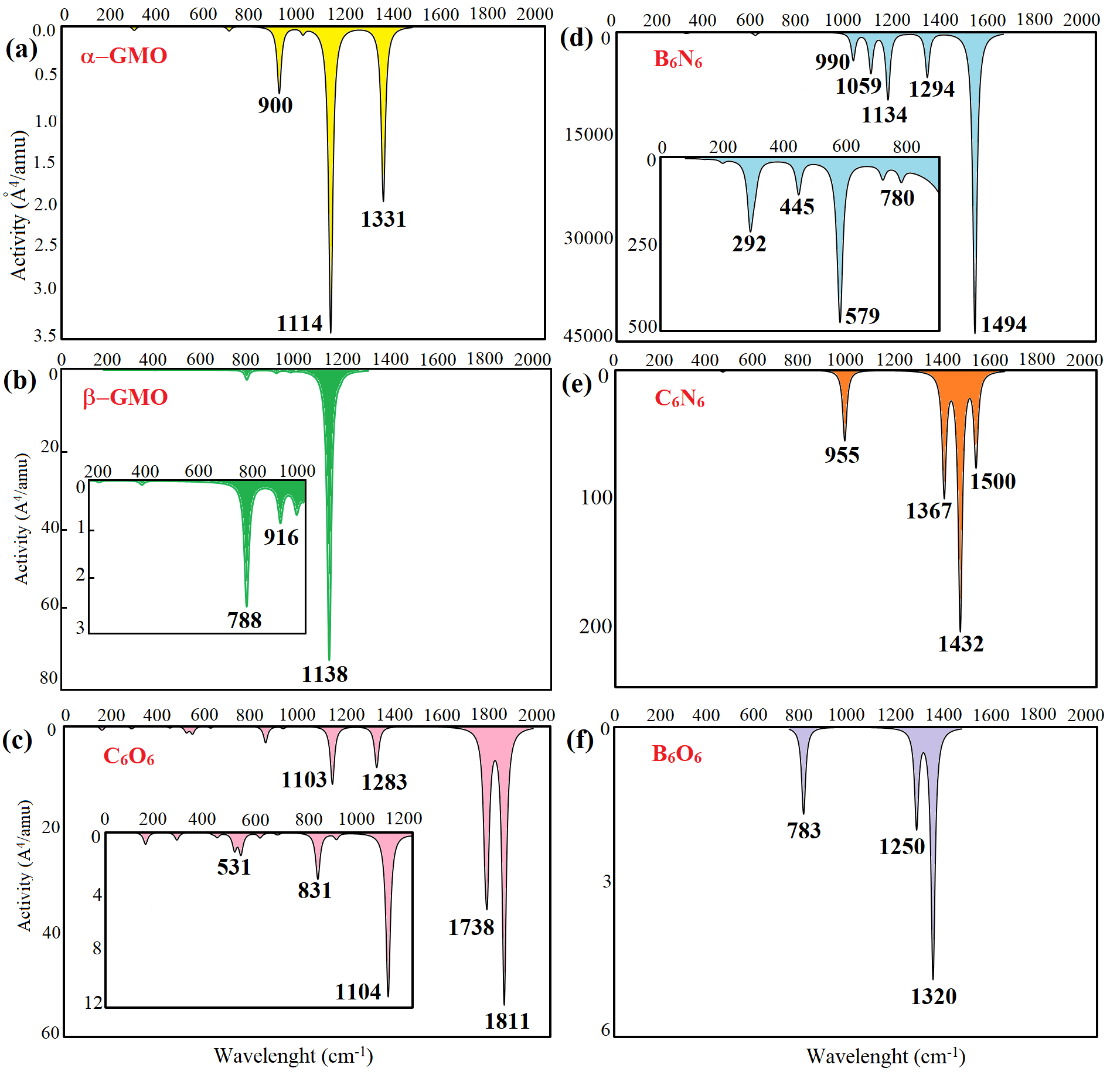}
\caption{Theoretical Raman spectra of (a) a-GMO, (b) b-GMO, (c) $C_6O_6$, (d) $B_6N_6$,
(e) $C_6N_6$, and (f) $B_6O_6$. The insets are zoomed in for the corresponding wavelengths.}
   \label{fig7}
\end{figure}

Fig.~\ref{fig7} displays the calculated Raman spectra of GMOs and COF-66. In GMOs and $C_6O_6$, the main characteristic Raman bands are located in the range of 700 cm$^{-1}$ to 1000 cm$^{-1}$ and have small peaks due to the C-C bonds ($\sim$ 800\,cm$^{-1}$) and phenyl ring ($\sim$ 1000\,cm$^{-1}$). The other bands found at 1100\,cm$^{-1}$ to 1300\,cm$^{-1}$ belong to C-O stretching\,\cite{nyquist2012handbook}.
In $C_6O_6$, two sharp peaks emerge at 1738\,cm$^{-1}$ and 1811\,cm$^{-1}$ which refer to the C=C and C=O stretching, respectively. The number of $\pi$ bonds in $C_6O_6$ is greater than in the other GMOs, and as a result, larger carrier mobility is expected.
The presence of holes in the structure actually blocks the carrier mobility across the surface. For $B_6N_6$ and $C_6N_6$, the sharp peak at 1492\,cm$^{-1}$ and 1432\,cm$^{-1}$, respectively, refers to the aromatic ring. The latter peak for $B_6N_6$ is the most characteristic of the boron nitride family. It results from the bond stretching of all pairs of sp$^2$ atoms and is analogous to the G peak in graphene\,\cite{griffin2018spectroscopic}. Additionally, in $C_6N_6$, a peak at 1500\,cm$^{-1}$ refers to the C=N bond.

The thermal evolution of the Gibbs free energy (G), enthalpy (H), and entropy (S) of a-GMO, b-GMO, and COF-66 were calculated, and the corresponding results are shown in Fig.~S5 of the Supplementary Information. For these structures, the trend of G, H, and S in the temperature range of 0 to 1000\,K is the same, and they all approach zero as the temperature approaches zero.
Furthermore, it was observed that the enthalpy increases, while the free energy decreases with increasing temperature.

In fact, as the temperature increases, the irregular thermal motion of the material increases, leading to an increase in both enthalpy and entropy, see Fig. S3. Fig.~S4(a) shows the temperature dependence of the Debye temperature ($\Theta$) for GMOs and COF-66, which is related to the highest frequency of lattice vibration. $\Theta$ can indicate the bond strength between atoms, and a larger $\Theta$ implies a higher melting point. It is observed that for temperatures lower than 500 K, $\Theta$ increases faster in comparison to the high-temperature regime ($>$500\, K). The heat capacity of a solid mainly originates from the thermal vibrations of the crystal lattice and the thermal movement of electrons. The contribution of the latter part is only significant at low temperatures and is negligible at high temperatures. Among COFs, $B_6N_6$ has the lowest $\Theta$. Additionally, there is a dip at about 25-95\,K, with the lowest value in the range of 415-1473\,K for COF-66.

The heat capacity (C$_p$) of GMOs and COF-66 as a function of temperature is shown in Fig.~S4(b) of the SI. It is evident that C$_p$ increases rapidly with a steep slope when T $<$ 500\,K, and increases slowly for T $>$ 500\,K. At high temperatures, C$_p$ approaches the Dulong$-$Petit classical limit\,\cite{luo2015electronic}. Additionally, C$_p$ rapidly approaches zero as the temperature approaches absolute zero. The C$_p$ of $C_6O_6$ is higher than that of the GMOs for temperatures below 500 K. Furthermore, for all temperatures, $B_6N_6$ exhibits the highest C$_p$ among the COF-66 structures.\\

\section{Hydrogen storage of $C_6B_6$}

\begin{figure}[tbp]
\includegraphics[width=0.90 \linewidth]{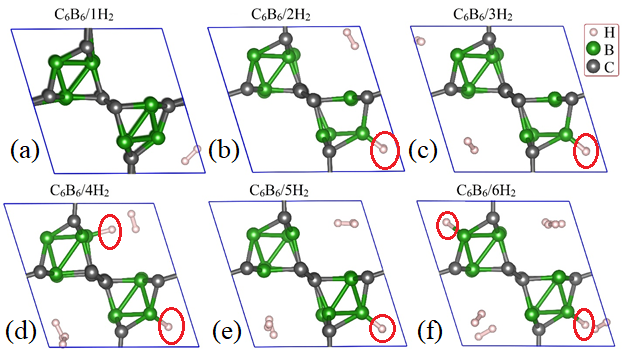}
\caption{ The sequential loading of hydrogen molecules over $C_6B_6$ up to six hydrogen molecules. Hydrogen molecule adsorption occurs via both dissociative and non-dissociative mechanisms.}
   \label{Fig8}
\end{figure}

 To explore the potential applications of COF-66, we conducted a study on the sequential loading of hydrogen molecules over $C_6B_6$, with a focus on up to six molecules, see Figu.~\ref{Fig8}. Our findings reveal that $H_2$ adsorption occurs via both dissociative and non-dissociative mechanisms. Moreover, we observed that as the number of hydrogen molecules increases, the binding nature of the pre-adsorbed hydrogens transitions from chemisorption to physisorption. In Fig.~\ref{Fig8}(a), a hydrogen molecule is physisorbed onto the surface. In a system containing two hydrogen molecules, when one hydrogen molecule dissociates, both separated hydrogen atoms can bond to boron atoms as indicated by the red circles in Fig.~\ref{Fig8}(b). For systems with more hydrogen molecules, further dissociation, and bonding can occur, as shown by the red circles in Figs.~\ref{Fig8}(d-f). However, it is important to note that a complete understanding of the hydrogen coverage of this system and its gravimetric storage capacity requires further research, which will be the subject of a separate paper.

Before concluding the paper, it would be worthwhile to mention that
{ one may consider synthesizing $C_6O_6$ using a similar method to that used for synthesizing $C_6N_6$ and $B_6O_6$ in Refs.\,\cite{cao2004synthesis,stredansky2018surface}. The ''oxalic acid condensation" method might be used, in which oxalic acid molecules condense to form a precursor and $C_6O_6$ is subsequently formed with the release of three water molecules. To achieve this, the oxalic acid must be purified by undergoing degassing stages at temperatures below its evaporation point ($>$470\,K).


\section{Concluding remarks}
In conclusion, this study utilized  extensive density functional theory calculations to predict new COF structures, named COF-66, with remarkable physical properties. We demonstrated that COF-66 (excluding $C_6B_6$) are porous 2D-crystalline solids with periodic voids and diverse electronic structures, making them ideal for guest-templated host networks to host active molecules/atoms in regular nano-arrays. Furthermore, we predicted a new 2D-crystalline oxocarbon, as well as porous 3D-$C_6B_6$ and 2D-$C_6B_6$ crystals, as members of a five-membered COF family with $X_6Y_6$ stoichiometry, where X $=$ B, C; Y $=$ B, N, O; and X $\neq$ Y. This newly introduced COF family, which includes already synthesized members such as $C_6N_6$ and $B_6O_6$, opens up new opportunities for the development of biosensors using COFs. Ultimately, this study  contributes to the expansion of COF research and its  potential applications.

\section{Acknowledgement}
This work was supported by the Iran National Science Foundation and Flemish
Science Foundation (FWO-Vl) and the Methusalem program.
N. Hassani expresses her gratitude to the Research Council of Razi University
for enabling her to conduct this research by providing essential resources and facilities.
M. Neek-Amal and N. Hassani acknowledge INSF support.



 \bibliography{references.bib}

\end{document}